\documentclass[12pt]{article}
\usepackage{amsmath,amssymb}

\makeatletter \@addtoreset{equation}{section}
\makeatother

\def\ben{\begin{equation}}
\def\een{\end{equation}}

  \let\n=\nu

\let\C=\Chi

\def\nn{\nonumber} \def\bd{\begin{document}} \def\ed{\end{document}}
\def\ds{\documentstyle} \let\fr=\frac \let\bl=\bigl \let\br=\bigr
\let\Br=\Bigr \let\Bl=\Bigl
\let\bm=\bibitem
\let\na=\nabla
\let\pa=\partial \let\ov=\overline
\newcommand{\be}{\begin{equation}}
\newcommand{\ee}{\end{equation}}
\def\ba{\begin{array}}
\def\ea{\end{array}}
\def\ft#1#2{{\textstyle{\frac{\scriptstyle #1}{\scriptstyle #2}}}}
\def\fft#1#2{\frac{#1}{#2}}
\def\del{\partial}
\def\vp{\varphi}
\def\sst#1{{\scriptscriptstyle #1}}
\def\oneone{\rlap 1\mkern4mu{\rm l}}
\def\td{\tilde}
\def\wtd{\widetilde}
\def\ie{\rm i.e.\ }
\def\dalemb#1#2{{\vbox{\hrule height .#2pt
        \hbox{\vrule width.#2pt height#1pt \kern#1pt
                \vrule width.#2pt}
        \hrule height.#2pt}}}
\def\square{\mathord{\dalemb{6.8}{7}\hbox{\hskip1pt}}}
\newcommand{\ho}[1]{$\, ^{#1}$}
\newcommand{\hoch}[1]{$\, ^{#1}$}
\newcommand{\bea}{\begin{eqnarray}}
\newcommand{\eea}{\end{eqnarray}}
\newcommand{\ra}{\rightarrow}
\newcommand{\lra}{\longrightarrow}
\newcommand{\Lra}{\Leftrightarrow}
\newcommand{\ap}{\alpha^\prime}
\newcommand{\bp}{\tilde \beta^\prime}
\newcommand{\tr}{{\rm tr} }
\newcommand{\Tr}{{\rm Tr} }
\def\0{{\sst{(0)}}}
\def\1{{\sst{(1)}}}
\def\2{{\sst{(2)}}}
\def\3{{\sst{(3)}}}
\def\4{{\sst{(4)}}}
\def\5{{\sst{(5)}}}
\def\6{{\sst{(6)}}}
\def\7{{\sst{(7)}}}
\def\8{{\sst{(8)}}}
\def\n{{\sst{(n)}}}
\def\cA{{{\cal A}}}
\def\cB{{{\cal B}}}
\def\cF{{{\cal F}}}
\def\tV{\widetilde V}
\def\tW{\widetilde W}
\def\tH{\widetilde H}
\def\tE{\widetilde E}
\def\tF{\widetilde F}
\def\tA{\widetilde A}
\def\im{{{\rm i}}}
\def\tY{{{\wtd Y}}}
\def\ep{{\epsilon}}
\def\vep{{\varepsilon}}
\def\R{\rlap{\rm I}\mkern3mu{\rm R}}
\def\bD{{{\bar D}}}

\def\R{\rlap{\rm I}\mkern3mu{\rm R}}
\def\bD{{{\bar D}}}
\def\R{{{\mathbb R}}}
\def\C{{{\mathbb C}}}
\def\H{{{\mathbb H}}}
\def\CP{{{\mathbb C}{\mathbb P}}}
\def\RP{{{\mathbb R}{\mathbb P}}}
\def\Z{{{\mathbb Z}}}
\def\bA{{{\mathbb A}}}
\def\bB{{{\mathbb B}}}
\def\bC{{{\mathbb C}}}
\def\bD{{{\mathbb D}}}
\def\bE{{{\mathbb E}}}
\def\bZ{{{\mathbb Z}}}
\def\Re{{{\frak{Re}}}}
\def\Im{{{\frak{Im}}}}
\def\cosec{{\,\hbox{cosec}\,}}
\def\Gm{{\Gamma_{\!\! -}}}
\def\Gp{{\Gamma_{\!\! +}}}
\def\stan{{standard }}
\def\nonstan{{supernumerary }}

\newcommand{\auth}{Z.-W. Chong}

\begin{document}
\begin{flushright}

MIFP-04-18\ \ \ \ \
{\bf hep-th/0411125}\\
Nov\  2004
\end{flushright}


\begin{center}

{\large {\bf General Metrics of $G_2$ and Spin(7)
Holonomy}}

\vspace{20pt}
\auth

\vspace{20pt} {\it George P. \& Cynthia W. Mitchell Institute for
Fundamental Physics,\\ Texas A\& M University, College Station, TX
77843-4242, USA} \vspace{30pt}

\underline{ABSTRACT}
\end{center}

Using a method introduced by Hitchin we obtain the system of first
order differential equations that determine the most general
cohomogeniety one $G_2$ holonomy metric with $S^3 \times S^3$
principal orbits.  The method is then applied to $G_2$ metric with
$S^3 \times T^3$ principal orbits in which an analytic solution is
obtained. The generalized metric has more free parameters than that
previously constructed. After showing that the generalization is
non-trivial a system of first order equations is obtained for new
Spin(7) metric with principal orbits $S^7$.

\pagebreak

\setcounter{page}{1}

\section{Introduction}

     Explicit metrics of special holonomy are of interest in both
physics and mathematics.  With the introduction of M-theory, seven and
eight dimensional manifolds with $G_2$ and Spin(7) holonomy become
particularly important since they provide natural candidates for
minimally supersymmetric compactifications.  (See for example,
\cite{Atiyahi,Atiyahii}.)  Explicit complete metrics for such compact
manifolds are unlikely, since they do not have continuous isometries.
However, for non-compact manifolds explicit metrics do exist, and
many $G_2$ and Spin(7) examples have been found
\cite{Hitchin,Bryant,GPP,Spin7i,%
0106026,Brandhuberi,Kannoi,Gukov,Kannoii,0112138,%
Brandhuberii,Chong,Cvetic}.

     The $G_2$ metrics with principal orbits $S^3 \times S^3$ are of
special interest because of their rich structure. The first non-singular 
example of
this kind was obtained in \cite{Bryant,GPP}, in which the two $S^3$ are
round.  
In \cite{0106026} a generalization of the metric ansatz depending on nine
unknown functions was given.  An ansatz with four unknown functions
was proposed in \cite{Brandhuberi}. More general metrics of this kind
were given in \cite{0112138,Brandhuberii}. For a review see
\cite{Cvetic}.
                                                                              
       The first example of a non-singular Spin(7) metric was given in
\cite{Bryant,GPP}, along with $G_2$ metrics including the one described 
above. The principal orbits of
this Spin(7) example are $S^7$, described as an $S^3$ bundle over $S^4$.  A
generalization of this metric was given in \cite{Spin7i}, by allowing
the $S^3$ fibres to be ``squashed". This generalization was shown to be
a special case in \cite{Gukov}, in which a new family of Spin(7)
metrics on a certain $\R^4$ bundle over $\CP^2$. For other
constructions see \cite{Kannoi,Kannoii}.

  In all the above examples, the metrics are of cohomogeneity one.
Hitchin \cite{Hitchin} gave a general practical tool for calculating
such special holonomy metrics of cohomogeneity one.  The examples
given there reproduce the Spin(7) metrics in \cite{Bryant,GPP,Spin7i}
and the $G_2$ metrics in \cite{Bryant,GPP,Brandhuberi}. This construction
was used in \cite{Chong} to obtain more general metrics of $G_2$
holonomy.  In this paper, we again employ this method to obtain
larger classes of metrics with $G_2$ and Spin(7) holonomy.

     The paper is organized as follows.  In section 2, We consider
the most general cohomogeneity one $G_2$ metric with $S^3\times S^3$
principal orbits.  We obtain first-order equations using the
Hitchin approach, which guarantees the existence of $G_2$
holonomy.  The metric is described by 18 functions satisfying 18
first-order differential equations, together with 7 consistent
algebraic constraints.  In section 3, we use the same approach to
study $G_2$ metrics with principal orbits $S^3 \times T^3$ \cite{Yau}, which
can be obtained by taking a contraction \cite{Chong} of one of the
$S^3$ factors to $T^3$.  We
obtain an analytic solution that has more non-trivial parameters
than those obtained in \cite{Yau}.  This demonstrates explicitly that our 
first-order system gives rise to a more general class of $G_2$ metrics than
any known previously.  In section 4, we apply this technique to
constructing a more general class of Spin(7) metrics whose principal
orbits are $S^7$.  We obtain a system of first-order equations,
which guarantees the existence of Spin(7) holonomy.  We conclude
our paper in section 5.

\section{General $G_2$ holonomy metric with $S^3\times S^3$
principal orbits}

     In this section we use the technique of \cite{Hitchin,Chong}
to obtain the most general cohomogeneity one $G_2$ metric with
$S^3\times S^3$ principal orbits.  A $G_2$ manifold is
characterised by its associative 3-form $\Phi_\3$, which has the
structure
\be
\Phi_\3 = dt\wedge \omega + \rho\,,\label{g23form}
\ee
where $\omega$ and $\rho$ are invariant 2-forms and 3-forms that do not
involve $dt$, satisfying the necessary condition
\be
\omega\wedge\rho=0\,.\label{omegarho}
\ee
Since $S^3$ is an $SU(2)$ group manifold, we can write the
vielbein for the $S^3\times S^3$ in terms of two sets of 
left-invariant $SU(2)$ 1-forms $\sigma_i$ and $\Sigma_i$, satisfying
\be
d\sigma_i=\fft12\epsilon_{ijk} \sigma_j\wedge \sigma_k\,,\qquad
d\Sigma_i=\fft12\epsilon_{ijk} \Sigma_j\wedge \Sigma_k\,.
\ee
We consider the most general 3-form $\rho$ 
 constructed from $\sigma_i$ and $\Sigma_i$
 \footnote{Note that here we break the anti-invariance 
\cite{Hitchin} under the $\mathbf {Z}/2$ that interchanges the two 
$S^3$ factors. 
This anti-invariance can be restored by setting $m=n$, $x_4=x_5$, $x_6=x_7$, 
and  $x_8=x_9$. It is similar for the following 4-form $\sigma$.}, given by
\bea
\rho&=&n\Sigma_1\Sigma_2\Sigma_3-m\sigma_1\sigma_2\sigma_3+x_1
d(\sigma_1\Sigma_1)+x_2d(\sigma_2\Sigma_2)+x_3d(\sigma_3\Sigma_3)\nn\\
&&+x_4d(\sigma_1\Sigma_2)
+x_5d(\sigma_2\Sigma_1)+x_6d(\sigma_2\Sigma_3)+x_7d(\sigma_3\Sigma_2)+
x_8d(\sigma_3\Sigma_1)+
x_9d(\sigma_1\Sigma_3)\,.
\eea
where $m$ and $n$ are constants, and $x_i$ are nine functions
depending on $t$.  In order to obtain $\omega$, we first consider the
most general 4-form $\sigma$, involving nine $t$-dependent
functions $y_i$:
\bea
\sigma&=&y_1\sigma_2\Sigma_2\sigma_3\Sigma_3+
y_2\sigma_3\Sigma_3\sigma_1\Sigma_1+y_3\sigma_1\Sigma_1
\sigma_2\Sigma_2+y_4\sigma_2\Sigma_3\sigma_3\Sigma_1+
y_5\sigma_3\Sigma_2\sigma_1\Sigma_3\nn\\
&&+y_6\sigma_3\Sigma_1\sigma_1\Sigma_2+
y_7\sigma_1\Sigma_3\sigma_2\Sigma_1+y_8\sigma_1\Sigma_2
\sigma_2\Sigma_3+y_9\sigma_2\Sigma_1\sigma_3\Sigma_2\,.
\label{g2sig}
\eea
Note that in this paper, in many complex equations when there is no
confusion, we shall drop the $\wedge$ notation for wedge products of
differential forms.  Following the approach of \cite{Hitchin}, we take
the ``square root'' of the 4-form, writing it as $\sigma=\ft12
\omega^2$.  Then we can write $\omega$ as
\be
\omega=a\sigma_1\Sigma_1+b\sigma_2\Sigma_2+
c\sigma_3\Sigma_3+e\sigma_2\Sigma_1+f\sigma_1\Sigma_2+
g\sigma_2\Sigma_3+h\sigma_3\Sigma_2+
j\sigma_3\Sigma_1+k\sigma_1\Sigma_3,
\ee
where
\bea
 a=\frac{y_2y_3-y_6y_7}{W},\quad b=\frac{y_1y_3-y_8y_9}{W},\quad 
c=\frac{y_1y_2-y_4y_5}{W},\nn\\
e=\frac{y_7y_9-y_3y_4}{W},\quad f=\frac{y_6y_8-y_3y_5}{W},\quad 
g=\frac{y_4y_8-y_1y_7}{W},\nn\\
h=\frac{y_5y_9-y_1y_6}{W},\quad j=\frac{y_4y_6-y_2y_9}{W},\quad 
k=\frac{y_5y_7-y_2y_8}{W}.
\eea 
The condition (\ref{omegarho}) now implies the algebraic constraints
\bea
-ax_4+bx_5-ex_2+fx_1-jx_7+hx_8=0,\!\!\!&&\!\!\!
-bx_6+cx_7-fx_9+gx_2-hx_3+kx_4=0,\nn\\
ax_9-cx_8+ex_6-gx_5+jx_3-kx_1=0,\!\!\!&&\!\!\!
-ax_5+bx_4+ex_1-fx_2+gx_9-kx_6=0,\nn\\
-bx_7+cx_6-ex_8-gx_3+hx_2+jx_5=0, \!\!\!&&\!\!\!
ax_8-cx_9+fx_7-hx_4-jx_1+kx_3=0.\label{g2con1}
\eea 

        Having obtained the ansatz for the associative 3-form $\Phi_\3$,
we can write down the metric for the $G_2$ manifold.  We define the
symmetric tensor density
\be 
B_{AB} = -\ft1{144} \Phi_{A C_1 C_2}\, \Phi_{B C_3 C_4}\,
\Phi_{C_5 C_6 C_7}\, \vep^{C_1\cdots C_7}\,,\label{dendef} 
\ee
where $\vep^{C_1\cdots C_7}$ is the Levi-Civita tensor density in
seven dimensions (with values $\pm1$ and 0). The metric tensor is then 
given by
\be
g_{AB} = \det(B)^{-1/9}\, B_{AB}\,.\label{metric}
\ee

    The Hamiltonian of the system can be written as $H=V(\rho) -
2W(\sigma)$, where $V(\rho)$ depends only on the tensor $\rho$, and
$W(\sigma)$ depends only on $\sigma$.  The function $V(\rho)$ is
defined by
\be
V(\rho) = \sqrt{-\ft16 K_a{}^b\, K_b{}_{\phantom{\Sigma}}^a}\,,
\ee
where
\be
K_a{}^b\equiv \ft1{12} \rho_{c_1 c_2 c_3}\, \rho_{c_4 c_5 a}\, 
\vep^{c_1 c_2 c_3 c_4 c_5 b}\,,
\ee
with $\vep^{c_1\cdots c_6}$ being the Levi-Civita 
tensor density in 6-dimensions.
The function $W(\sigma)$ is calculated from
\be
W(\sigma)^2 = \ft1{48}\, \vep_{c_1\cdots c_6}\, \td\sigma^{c_1 c_2}\,
\td\sigma^{c_3c_4}\,\td\sigma^{c_5c_6}\,,
\ee
where
\be
\td \sigma^{ab} \equiv \ft1{24}\, \vep^{ab c_1c_2c_3c_4}\, 
\sigma_{c_1c_2c_3c_4}\,.
\ee
For our specific example, we find that 
\bea
V&\equiv&\sqrt{-U}\,,\nn\\
U&=&m^2n^2-2mn\sum_{i=1}^{9}x_{i}^2+\sum_{i=1}^{9}x_{i}^{4}\nn\\
&&-4(m+n)(x_1x_2x_3-x_3x_4x_5-x_1x_6x_7+x_4x_6x_8+x_5x_7x_9-x_2x_8x_9)\nn\\
&&-2x_{1}^{2}(x_{2}^{2}+x_{3}^{2}-x_{4}^{2}-x_{5}^{2}+x_{6}^{2}+x_{7}^{2}
-x_{8}^{2}-x_{9}^{2})\nn\\
&&+2x_{2}^{2}(-x_{3}^{2}+x_{4}^{2}+x_{5}^{2}+x_{6}^{2}+
x_{7}^{2}-x_{8}^{2}-x_{9}^{2})+
2x_{3}^{2}(-x_{4}^{2}-x_{5}^{2}+x_{6}^{2}+x_{7}^{2}+x_{8}^{2}+x_{9}^{2})\nn\\
&&+2x_{4}^{2}(-x_{5}^{2}-x_{6}^{2}+x_{7}^{2}-x_{8}^{2}+
x_{9}^{2})+2x_{5}^{2}(x_{6}^{2}-x_{7}^{2}
+x_{8}^{2}-x_{9}^{2})\nn\\
&&+2x_{6}^{2}(-x_{7}^{2}-x_{8}^{2}+x_{9}^{2})+2x_{7}^{2}(x_{8}^{2}-
x_{9}^{2})+2x_{8}^{2}(-x_{9}^{2})\nn\\
&&+8x_1(x_2x_4x_5+x_4x_7x_8+x_5x_6x_9+x_3x_8x_9)+8x_3x_5x_6x_8\nn\\
&&+8x_2x_4x_6x_9+8x_3x_4x_7x_9+8x_2x_3x_6x_7+8x_2x_5x_7x_8\,,\nn\\
W&=&(y_1y_2y_3+y_4y_6y_8+y_5y_7y_9-y_3y_4y_5-y_2y_8y_9-y_1y_6y_7)^{\fft12}\,.
\eea

      The manifold with $G_2$ holonomy is then governed by a set of
first-order differential equations following from the Hamiltonian
flow \cite{Hitchin}
\be 
\dot x_{i}=-\frac{\partial H}{\partial y_i},\quad \dot
y_i=\frac{\partial H}{\partial x_i},\label{hamflow}
\ee
where the dot denotes a derivative with respect to the ``time''
variable $t$, together with the Hamiltonian constraint $H=0$.  Thus we
have
\bea
\dot x_1=\frac{y_2y_3-y_6y_7}{W},\quad \dot
x_2=\frac{y_1y_3-y_8y_9}{W},\quad \dot
x_3=\frac{y_1y_2-y_4y_5}{W}, \nn\\ \dot
x_4=\frac{y_6y_8-y_3y_5}{W},\quad \dot
x_5=\frac{y_7y_9-y_3y_4}{W},\quad
\dot x_6=\frac{y_4y_8-y_1y_7}{W},\nn\\
\dot x_7=\frac{y_5y_9-y_1y_6}{W},\quad \dot
x_8=\frac{y_4y_6-y_2y_9}{W},\quad \dot x_9=
\frac{y_5y_7-y_2y_8}{W}.\eea \bea \dot
y_1&=&[mnx_1+(m+n)(x_2x_3-x_6x_7)+x_1(x_{2}^{2}+x_{3}^{2}-
x_{1}^{2}-x_{4}^{2}-x_{5}^{2}+x_{6}^{2}
+x_{7}^{2}-x_{8}^{2}-x_{9}^{2})\nn\\
&&-2(x_2x_4x_5+x_4x_7x_8+x_5x_6x_9+x_3x_8x_9)]/W,\nn\\
\dot
y_2&=&[mnx_2+(m+n)(x_3x_1-x_8x_9)+x_2(x_{3}^{2}+x_{1}^{2}-
x_{2}^{2}-x_{4}^{2}-x_{5}^{2}-x_{6}^{2}
-x_{7}^{2}+x_{8}^{2}+x_{9}^{2})\nn\\
&&-2(x_1x_4x_5+x_4x_6x_9+x_5x_7x_8+x_3x_6x_7)]/W,\nn\\
\dot
y_3&=&[mnx_3+(m+n)(x_1x_2-x_4x_5)+x_3(x_{1}^{2}+x_{2}^{2}-
x_{3}^{2}+x_{4}^{2}+x_{5}^{2}-x_{6}^{2}
-x_{7}^{2}-x_{8}^{2}-x_{9}^{2})\nn\\
&&-2(x_1x_8x_9+x_5x_6x_8+x_4x_7x_9+x_2x_6x_7)]/W,\nn\eea 
\bea
\dot
y_4&=&[mnx_4+(m+n)(x_6x_8-x_3x_5)+x_4(-x_{1}^{2}-x_{2}^{2}+
x_{3}^{2}-x_{4}^{2}+x_{5}^{2}+x_{6}^{2}
-x_{7}^{2}+x_{8}^{2}-x_{9}^{2})\nn\\
&&-2(x_1x_2x_5+x_1x_7x_8+x_2x_6x_9+x_3x_7x_9)]/W,\nn \\
\dot
y_5&=&[mnx_5+(m+n)(x_7x_9-x_3x_4)+x_5(-x_{1}^{2}-x_{2}^{2}+
x_{3}^{2}+x_{4}^{2}-x_{5}^{2}-x_{6}^{2}
+x_{7}^{2}-x_{8}^{2}+x_{9}^{2})\nn\\
&&-2(x_1x_2x_4+x_1x_6x_9+x_3x_6x_8+x_2x_7x_8)]/W,\nn \\
\dot
y_6&=&[mnx_6+(m+n)(x_4x_8-x_1x_7)+x_6(x_{1}^{2}-x_{2}^{2}-
x_{3}^{2}+x_{4}^{2}-x_{5}^{2}-x_{6}^{2}
+x_{7}^{2}+x_{8}^{2}-x_{9}^{2})\nn\\
&&-2(x_1x_5x_9+x_3x_5x_8+x_2x_4x_9+x_2x_3x_7)]/W,\nn\\
\dot{y_7}&=&[mnx_7+(m+n)(x_5x_9-x_1x_6)+x_7(x_{1}^{2}-x_{2}^{2}-
x_{3}^{2}-x_{4}^{2}+x_{5}^{2}+x_{6}^{2}
-x_{7}^{2}-x_{8}^{2}+x_{9}^{2})\nn\\
&&-2(x_1x_4x_8+x_2x_3x_6+x_3x_4x_9+x_2x_5x_8)]/W,\nn\\
\dot
y_8&=&[mnx_8+(m+n)(x_4x_6-x_2x_9)+x_8(-x_{1}^{2}+x_{2}^{2}-
x_{3}^{2}+x_{4}^{2}-x_{5}^{2}+x_{6}^{2}
-x_{7}^{2}-x_{8}^{2}+x_{9}^{2})\nn\\
&&-2(x_1x_4x_7+x_1x_3x_9+x_3x_5x_6+x_2x_5x_7)]/W,\nn\\
\dot
y_9&=&[mnx_9+(m+n)(x_5x_7-x_2x_8)+x_9(-x_{1}^{2}+x_{2}^{2}-
x_{3}^{2}-x_{4}^{2}+x_{5}^{2}-x_{6}^{2}
+x_{7}^{2}+x_{8}^{2}-x_{9}^{2})\nn\\
&&-2(x_1x_5x_6+x_1x_3x_8+x_2x_4x_6+x_3x_4x_7)]/W\,. 
\eea 
The Hamiltonian constraint implies that
\be
U=-4\,(y_1y_2y_3+y_4y_6y_8+y_5y_7y_9-y_3y_4y_5-
y_2y_8y_9-y_1y_6y_7)\,,\label{hamilton}
\ee

      Finally we present the explicit form of the metric, which is given by
\bea
ds^2&=&dt^2+g_{11}\sigma_{1}^2+2g_{12}\sigma_{1}\sigma_{2}+
2g_{13}\sigma_{1}\sigma_{3}+2g_{14}\sigma_{1}\Sigma_{1}+
2g_{15}\sigma_{1}\Sigma_{2}+2g_{16}\sigma_{1}\Sigma_{3}\nn\\
&&+g_{22}\sigma_{2}^2+2g_{23}\sigma_{2}\sigma_{3}+
2g_{24}\sigma_{2}\Sigma_{1}+2g_{25}\sigma_{2}\Sigma_{2}+
2g_{26}\sigma_{2}\Sigma_{3}\nn\\
&&g_{33}\sigma_{3}^2+2g_{34}\sigma_{3}\Sigma_{1}+
2g_{35}\sigma_{3}\Sigma_{2}+2g_{36}\sigma_{3}\Sigma_{3}\nn\\
&&g_{44}\Sigma_{1}^2+2g_{45}\Sigma_{1}\Sigma_{2}+
2g_{46}\Sigma_{1}\Sigma_{3}+g_{55}\Sigma_{2}^2+
2g_{56}\Sigma_{2}\Sigma_{3}+g_{66}\Sigma_{3}^2\,,
\eea 
where $g_{ij}$ can be calculated in a straightforward way from
(\ref{metric}).  Owing to the complexity of the structures, we shall
not present the explicit results here.  We did verify that the system
of first-order equations does imply the closure and co-closure of the
associative 3-form, which demonstrates that the metric indeed has
holonomy $G_2$.

    By the above construction we have obtained $G_2$ metrics involving 
18 functions $x_i$
and $y_i$, and two constants $m$ and $n$, governed by 7 algebraic
equations (\ref{g2con1},\ref{hamilton}), and $(18-7)=11$ independent
first-order equations.


\section{$G_2$ holonomy metric with $S^{3}\times T^{3}$ principal orbits}

   The $SU(2)$ group associated with an $S^3$ can be contracted in three
different ways, namely the Euclidean, Heisenberg, and Abelian contractions
(see, for example, \cite{Chong}).  Here we consider the Abelian contraction 
for $\sigma_i$.  To do this, we define $\sigma_i=\lambda\,\alpha_i$, and then
send $\lambda \rightarrow 0$.  Thus we have $d\alpha_i=0$, and
correspondingly the $S^3$ becomes (locally) $T^3$.

  We start with the 3-form $\rho$ and 4-form $\sigma$ 
\bea
\rho&=&n\Sigma_{1}\Sigma_{2}\Sigma_{3}-m\alpha_{1}\alpha_{2}\alpha_{3}+x_{1}d(\Sigma_{1}\alpha_{1})+x_{2}d(\Sigma_{2}\alpha_{2})+x_{3}d(\Sigma_{3}\alpha_{3})\nn\\
&&+x_{4}d(\Sigma_{1}\alpha_{2})+x_{5}d(\Sigma_{2}\alpha_{1}),\label{3formrho}\\
\sigma&=&y_{1}\Sigma_{2}\alpha_{2}\Sigma_{3}\alpha_{3}+y_{2}\Sigma_{3}\alpha_{3}\Sigma_{1}\alpha_{1}+y_{3}\Sigma_{1}\alpha_{1}\Sigma_{2}\alpha_{2}\nn\\
&&+y_{4}\Sigma_{2}\alpha_{3}\Sigma_{3}\alpha_{1}+y_{5}\Sigma_{3}\alpha_{2}\Sigma_{1}\alpha_{3}\label{sigma}
\eea 
Note that a 3-form $\rho$ without the $x_4$ and $x_5$ terms, and 
correspondingly a 4-form $\sigma$ without the $y_4$ and $y_5$ 
terms, were considered in \cite{Yau}. We will see later that 
the more general 3-form $\rho$ and 4-form $\sigma$ considered here will 
give rise to an off-diagonal term in the metric.

The Hamiltonian is given by 
\bea 
H=V-2W=\sqrt{-U}-2W 
\eea
where
\bea
U=m^2 n^2 +4m(x_{1}x_{2}x_{3}-x_{3}x_{4}x_{5}),\ \ \ 
W=(y_{1}y_{2}y_{3}-y_{3}y_{4}y_{5})^{\frac{1}{2}}. 
\eea 
The co-associative 3-form is $\Phi_{(3)}=dt\wedge\omega+\rho$, where
\begin{equation}
\omega=\frac{y_{2}y_{3}}{W}\Sigma_{1}\alpha_{1}+\frac{y_{3}y_{1}}{W}\Sigma_{2}\alpha_{2}+\frac{W}{y_{3}}\Sigma_{3}\alpha_{3}-\frac{y_{3}y_{4}}{W}\Sigma_{2}\alpha_{1}-\frac{y_{3}y_{5}}{W}\Sigma_{1}\alpha_{2}.
\end{equation}
A $G_2$ holonomy metric is obtained if $x_i$ and $y_i$ satisfy
the Hamiltonian flow equation
\bea \dot{x_i}=-\frac{\partial
H}{\partial y_{i}}, \quad \dot{y_i}=\frac{\partial H}{\partial
x_{i}}\,,
\eea 
which results in 
\bea
\dot{x_1}&=&\frac{y_{2}y_{3}}{W},\quad \ \ \ \ \
\dot{x_2}=\frac{y_{3}y_{1}}{W},\quad \ \ \ \
\dot{x_3}=\frac{y_{1}y_{2}-y_{4}y_{5}}{W},
\nn\\
\dot{x_4}&=&-\frac{y_{3}y_{5}}{W},\quad \quad \dot{x_5}=-\frac{y_{3}y_{4}}{W},\nn\\
\dot{y_1}&=&\frac{2mx_{2}x_{3}}{\sqrt{-U}},\quad \
\dot{y_2}=\frac{2mx_{3}x_{1}}{\sqrt{-U}},\quad\dot{y_3}=\frac{2m(x_{1}x_{2}-x_{4}x_{5})}{\sqrt{-U}},\quad
\nn\\
\dot{y_4}&=&-\frac{2mx_{3}x_{5}}{\sqrt{-U}},\ \
\dot{y_5}=-\frac{2mx_{3}x_{4}}{\sqrt{-U}}. 
\eea 
A simpler system of equations can be obtained by a change of variable from
$t$ to $\td t$ \cite{Yau},
\bea 
\frac{dt}{d\tilde{t}}=4\sqrt{y_{1}y_{2}y_{3}-y_{3}y_{4}y_{5}}, \label{tchange}
\eea 
which is equivalent to considering the Hamiltonian flow 
\bea
\tilde{H}=-m^2n^2+4m(x_{1}x_{2}x_{3}-x_{3}x_{4}x_{5})-4(y_{1}y_{2}y_{3}-y_{3}y_{4}y_{5})=-m^2n^2-4mX-4Y
\eea
with 
\bea X=x_{1}x_{2}x_{3}-x_{3}x_{4}x_{5},
Y=y_{1}y_{2}y_{3}-y_{3}y_{4}y_{5}\,.
\eea 
The flow equation becomes 
\bea
x_1'&=&4y_{2}y_{3},\  \ \ \  \ \ x_2'=4y_{3}y_{1},\quad \quad x_3'=4(y_{1}y_{2}-y_{4}y_{5}),\nn\\
x_4'&=&-4y_{3}y_{5},\  \ \ \ x_5'=-4y_{4}y_{3},\quad\nn\nn\\
y_1'&=&4mx_{2}x_{3},\  \ \ y_2'=4mx_{3}x_{1},\quad y_3'=4m(x_{1}x_{2}-x_{4}x_{5}),\nn\\
y_4'&=&-4mx_{3}x_{5},\quad y_5'=-4mx_{3}x_{4}. 
\eea 
where the prime denotes a derivative with respect to $\td t$.
In addition $x_i$ and $y_i$ satisfy the constraint from the
requirement $\omega\wedge\rho=0$, namely
\bea
y_{2}x_{4}-y_{1}x_{5}-y_{4}x_{2}+y_{5}x_{1}=0. 
\eea 
Suggested by \cite{Yau}  we find the following conserved quantities 
\bea x_2
y_2 -x_1 y_1 =k_1 ,\quad
x_5 y_5 -x_4 y_4 &=&k_2 ,\quad x_1 y_1 +x_4 y_4 -x_3 y_3 =k_3\nn\\
x_5 y_1 +x_2 y_4 &=&x_1 y_5+x_4 y_2 =\lambda\,, 
\eea 
where $k_1$, $k_2$, $k_3$ and $\lambda$ are constants. Defining 
$z_3 =x_3 y_3$, we find that 
\bea
\frac{dz_3}{d\tilde{t}}&=&4Y-4mX=m^2n^2+8Y\\
\frac{d^2z_3}{d\tilde{t}^2}&=&8\frac{dY}{d\tilde{t}}=-32m[3z_{3}^{2}+2(k_1
+k_2+ 2k_3)z_3 +k_{3}(k_1 +k_2 +k_3 )-\lambda^2]\,. 
\eea 
This can be integrated explicitly, and $z_3$ can be written in terms 
of Weierstrass
function, which has a second order pole. Near a pole
$x_{i}(\tilde{t})$ and $y_{i}(\tilde{t})$ takes the approximate form 
$x_{i}\sim
y_{i}\sim \frac{1}{\tilde{t}}$ \cite{Yau}. Written in terms of $t$ through
the relation (\ref{tchange}), we have 
\bea x_1&=&A_1
t^2,\quad y_1 =4A_2A_3t^2,\quad \nn\\
x_2&=&A_2 t^2,\quad y_2 =4A_3A_1 t^2,\quad \nn\\
x_3&=&A_3 t^2,\quad y_3 =4(A_1A_2-A_4A_5) t^2,\quad \nn\\
x_4&=&A_4 t^2,\quad y_4 =-4A_3A_5 t^2,\quad \nn\\ x_5&=&A_5
t^2,\quad y_5 =-4A_3A_4t^2.\quad 
\eea 
If we set $A_1 A_2 A_3 -A_3A_4 A_5 =\frac{m}{64}$, a $G_2$ metric of topology $\R^4 \times
T^3$ can be obtained.
\begin{equation}
ds^2=dt^2+\frac{1}{4}t^2(\Sigma_{1}^2+\Sigma_{2}^2+\Sigma_{3}^2)+16[(A_{1}^2-A_{5}^2)\alpha_{1}^2+2(A_1
A_4 +A_2 A_5)\alpha_1 \alpha_2
+(A_{2}^2-A_{4}^2)\alpha_{2}^2+A_{3}^2\alpha_{3}^2]
\end{equation}
In the construction of \cite{Yau}, the $G_2$ metric had two free 
parameters.  Here, however,
the generalized metric has four independent parameters. Also we note that 
there is an off-diagonal term in the metric which results from the 
extra $x_4$ and $x_5$ terms in the 3-form (\ref{3formrho}). This shows that
the metric considered here is more general, though it has the same
topology as that in \cite{Yau}.

\section{Spin(7) metric with principle orbit $S^7$}
In this section we will consider new Spin(7) metric with principal 
orbits $S^7$ described as an $SU(2)$ bundle over $S^4$. The three 
invariant $SU(2)$ connection 1-forms are $\alpha_i$ ($i$=1,2,3), and 
$\omega_i$ ($i$=1,2,3) are the curvature 2-forms. The structure equations 
for the principal orbits are 
\bea d\Sigma_1&=&\omega_1 -2\Sigma_2 \Sigma_3 ,\quad \quad \ \ d\Sigma_2
=\omega_2 -2 \Sigma_3 \Sigma_1, \quad \quad \ \ \ d\Sigma_3
=\omega_3 -2 \Sigma_1
\Sigma_2,\nn\\
d\omega_1 &=&2(\omega_2 \Sigma_3 -\omega_3 \Sigma_2 ),\quad
d\omega_2 =2(\omega_3 \Sigma_1 -\omega_1 \Sigma_3 ),\quad
d\omega_3 =2(\omega_1 \Sigma_2 -\omega_2 \Sigma_1 ) 
\eea 
with
\bea \omega_1 =-(\Sigma_0 \Sigma_1 +\Sigma_2 \Sigma_3 ),\quad
\omega_2 =-(\Sigma_0 \Sigma_2 +\Sigma_3 \Sigma_1 ),\quad \omega_3
=-(\Sigma_0 \Sigma_3 +\Sigma_1 \Sigma_2 )\,, 
\eea 
where the basis $\Sigma_\mu$ ($\mu$=0,1,2,3) give the 
standard metric $ds^2=\Sigma_0^2+\Sigma_1^2+\Sigma_2^2+\Sigma_3^2$ on $S^4$. 

We consider the following exact 4-form constructed from $\alpha_i$ and $\omega_i$
\bea \rho&=&x_1 d(\alpha_1 \omega_1)+x_2 d(\alpha_2
\omega_2)+x_3 d(\alpha_3 \omega_3)+x_4 d(2\alpha_1 \alpha_2
\alpha_3)+x_5 d(\alpha_1 \omega_2)\label{rho}\\ &=&2(x_1 +x_2
+x_3)\Sigma_0 \Sigma_1 \Sigma_2 \Sigma_3-2(-x_1 +x_2 +x_3
+x_4)\alpha_2 \alpha_3 (\Sigma_0 \Sigma_1 +\Sigma_2 \Sigma_3)\nn\\
&&-2(x_1 -x_2 +x_3 +x_4)\alpha_3 \alpha_1 (\Sigma_0 \Sigma_2
+\Sigma_3 \Sigma_1)\nn\\
&&-2(x_1 +x_2 -x_3 +x_4)\alpha_1 \alpha_2 (\Sigma_0 \Sigma_3
+\Sigma_1 \Sigma_2)\nn\\ &&+2x_5 [\alpha_2 \alpha_3 (\Sigma_0
\Sigma_2 +\Sigma_3 \Sigma_1)+\alpha_3 \alpha_1 (\Sigma_0 \Sigma_1
+\Sigma_2 \Sigma_3)]\label{4formrho}.
\eea 
Note that the above 4-form $\rho$ but without the $x_5$ term was
considered in \cite{Hitchin}, reproducing the Spin(7) metrics obtained
in \cite{Spin7i}. Also note that the most general 4-form constructed
from $\alpha_i$ and $\Sigma_\mu$ was written down in \cite{Kannoiii}.
In order to follow Hitchin's procedure, one should then calculate the
metric that is implied by the choice of 4-form, and solve the
equations following from the Hamiltonian flow.  In \cite{Kannoiii} the
form of the metric was instead imposed as an additional constraint,
and in fact this was more restrictive than was implied by the choice
of 4-form, leading to a highly constrained solution set.  In this
paper, by contrast, we consider a more modest generalization of
previous choices for the 4-form ansatz (by including the $x_5$ term in
(\ref{4formrho})), but we do follow the Hitchin procedure and {\it
derive} the form of the metric, rather than imposing it as an
additional ansatz.  We shall see that, as in Section 3, the extra
$x_5$ term will give rise to an off-diagonal term in the Spin(7)
metric. (Such terms were not included in the metric ansatz considered
in \cite{Kannoiii}.)  In consequence, we obtain new Spin(7) metrics
that were not found in the analysis in \cite{Kannoiii}.  The extension 
to most general 4-form ansatz considered in \cite{Kannoiii}, and
with the metric derived from this ansatz, gives a more complicated system of 
first-order equations, analogous to those obtained for $G_2$ metrics in
section 2.  We shall not present these here, since they are rather 
involved but straightforward to derive.

    To calculate the metric we first construct the dual tensor density 
\bea
\tilde{\rho}^{abc}=\frac{1}{4!}\varepsilon^{abcd_1 d_2 d_3
d_4}\rho_{d_1 d_2 d_3 d_4}.
\eea 
Then, by defining the symmetric tensor density
\bea
H^{ab}=-\frac{1}{144}\tilde{\rho}^{ac_1 c_2}\tilde{\rho}^{bc_3
c_4}\tilde{\rho}^{c_5 c_6 c_7}\varepsilon_{c_1 c_2 \ldots
c_7}\,,
\eea
we can calculate the volume from $V=|\det H|^{1/12}$, finding
\bea 
V=a^2(a^4b_{1}^2b_{2}^2b_{3}^2-b_1 b_2 v_{1}^2 )^{\fft12}\,,
\eea
where
\bea
a^4&=&2(x_1 +x_2 +x_3),\quad \quad  a^2b_2 b_3=2(-x_1 +x_2 +x_3
+x_4),\nn\\ 
a^2 b_3 b_1&=&2(x_1 -x_2 +x_3 +x_4), a^2b_1 b_2 =2(x_1
+x_2 -x_3 +x_4), \ \ v_1 =-2x_5\label{abx}. 
\eea 
The gradient flow equation is given by 
\bea \frac{\partial V}{\partial
x_1}&=&2(-\dot{x_1}+\dot{x_2}+\dot{x_3}+\dot{x_4}),\quad\frac{\partial
V}{\partial x_2}=2(\dot{x_1}-\dot{x_2}+\dot{x_3}+\dot{x_4}),\nn\\
\frac{\partial V}{\partial
x_3}&=&2(\dot{x_3}+\dot{x_2}-\dot{x_3}+\dot{x_4}),\ \
\quad\frac{\partial
V}{\partial x_4}=2(\dot{x_1}+\dot{x_2}+\dot{x_3}),\nn\\
\frac{\partial V}{\partial x_5}&=&-2\dot{x_5}.
\eea 
It can be rewritten as 
\bea
\frac{da^4}{dt}&=&\frac{a^6}{V}b_1 b_2 b_3 (b_1 +b_2 +
b_3)-\frac{a^2v_{1}^2}{V},\nn\\ 
\frac{da^2b_2
b_3}{dt}&=&\frac{V}{2a^4}+\frac{a^6}{V}b_1 b_2 b_3 (-b_1 +b_2
+b_3)-\frac{a^2v_{1}^2}{V},\nn\\ 
\frac{da^2b_3
b_1}{dt}&=&\frac{V}{2a^4}+\frac{a^6}{V}b_1 b_2 b_3 (b_1 -b_2
+b_3)-\frac{a^2v_{1}^2}{V},\nn\\ 
\frac{da^2b_1
b_2}{dt}&=&\frac{V}{2a^4}+\frac{a^6}{V}b_1 b_2 b_3 (b_1 +b_2
-b_3)-\frac{a^2v_{1}^2}{V},\nn\\
\frac{dv_1}{dt}&=&2\frac{a^4b_1 b_2}{V}v_1.\label{Spin7ODE} 
\eea
The metric is obtained from
\bea
h_{ab}=|\det H|^{1/6}H_{ab} 
\eea
and it can be written as 
\bea
ds^2=dt^2+h_{11}\sigma_{1}^2+h_{22}\sigma_{2}^2+h_{33}\sigma_{3}^2+
2h_{12}\sigma_{1}\sigma_{2}+h_{\Sigma}^{2}(\Sigma_{1}^2+
\Sigma_{2}^2+\Sigma_{3}^2+\Sigma_{4}^2)\,,
\label{Spin7Metric}
\eea 
where 
\bea h_{11}&=&\frac{b_1 b_2
(a^4b_{1}^2b_{3}^2+v_{1}^2)}{a^4 b_1 b_2 b_{3}^2-v_{1}^2},\quad \
h_{22}=\frac{b_1 b_2 (a^4b_{2}^2b_{3}^2+v_{1}^2)}{a^4 b_1 b_2
b_{3}^2-v_{1}^2},\quad h_{33}=b_{3}^2-\frac{v_{1}^2}{a^4b_1
b_2}\nn\\ h_{12}&=&\frac{a^2b_1 b_2 (b_1 +b_2)b_3 v_1}{a^4 b_1 b_2
b_{3}^2-v_{1}^2},\quad h_{\Sigma}=a^2. 
\eea 
This metric  together with the system of first-order 
equations (\ref{Spin7ODE}) guarantees  
\bea \frac{\partial
\rho}{\partial t}=d(*\rho).
\eea 
which shows that the metric determined by (\ref{Spin7ODE}) has Spin(7)
holonomy. We see that $h_{12}$ gives rise to an off-disgonal 
term in (\ref{Spin7Metric}), which is proportional to $x_5$. Through the 
relation in (\ref{abx}), this off-diagonal term results from the extra 
$x_5$ term  in (\ref{4formrho}).

\section{Discussion}

In this paper we have obtained a class of cohomogeneity one
$G_2$ metrics with $S^3\times S^3$ principal orbits,  derived from
 the most general invariant 3-forms $\rho$ and
4-forms $\sigma$ which break the  anti-invariance under the $\Z/2$ action
interchanging the two $S^3$ factors. This should be the most general 
cohomogeniety one
$G_2$ metric that one can obtain with principal orbits $S^3 \times
S^3$. Although we cannot obtain the explicit solution to the highly coupled
system of non-linear first-order differential equations, we luckily obtained
an analytic solution to the
metric with principal orbits $S^3 \times T^3$, which is arises as
a group contraction \cite{Chong} of one of the two $S^3$ factors. This 
type of contraction was considered in \cite{Yau}. By showing that
our solution has more free parameters than the solutions obtained in
\cite{Yau}, we can conclude that this
generalization is a non-trivial one.

   We also considered Spin(7) metrics with principal orbits $S^7$,
with a similar generalization of previous results obtained by writing
down a more general exact 4-form, constructed from the
$SU(2)$-connection 1-form $\alpha_i$ and curvature 2-form $\omega_i$.
It was shown there that the most general exact 4-forms one can
consider has ten parameters. In this paper we only considered a five
parameter exact 4-form, which is already sufficient to imply the new
feature of off-diagonal elements in the Spin(7) metrics.  We
concentrated on how the generalization modified the system of
differential equations which determined the Spin(7) metrics. We also
showed how the metrics looks. The formalism presented here can also be
applied in a straightforward way to Spin(7) metrics whose principal
orbits are Aloff-Wallach spaces, i.e.  $SU(3)/U(1)$ cosets.

\section*{Acknowledgments}
The author is grateful to Gary Gibbons, Hong L\"u and Chris Pope for useful discussions. He is supported in
part by DOE grant DE-FG03-95ER40917.

\end{document}